\documentclass{article}
\usepackage{arxiv}
\usepackage{orcidlink}
\usepackage{authblk}
\usepackage{amsmath}
\usepackage{amssymb}
\usepackage{appendix}
\usepackage{fancyhdr}
\usepackage{booktabs}       
\usepackage{float}

\usepackage{hyperref}
\hypersetup{
    colorlinks=true,
    linkcolor=blue,
    citecolor=blue
}

\usepackage{algorithm}
\usepackage{algorithmicx}
\usepackage{algpseudocode}
\usepackage{tabularx}
\makeatletter
\newcommand{\multiline}[1]{%
  \begin{tabularx}{\dimexpr\linewidth-\ALG@thistlm}[t]{@{}X@{}}
    #1
  \end{tabularx}
}
\makeatother

\begin{document}


\title{Bayesian Profile Regression using Variational Inference to Identify Clusters of Multiple Long-Term Conditions Conditioning on Mortality in Population-Scale Data}
\fancyhf{}
\fancyhead[RO]{BPR using VI for clustering MLTC}
\fancyhead[LO]{\thepage}
\author[1,$\ast$]{James Rafferty\orcidlink{0000-0002-1667-7265}}
\author[2]{Keith R. Abrams\orcidlink{0000-0002-7557-1567}}
\author[3]{Munir Pirmohamed\orcidlink{0000-0002-7534-7266}}
\author[4]{Mark Davies\orcidlink{0000-0002-0688-3129}}
\author[1]{Rhiannon K. Owen\orcidlink{0000-0001-5977-376X}}


\affil[1]{Health Data Research UK, Swansea University Medical School, Swansea University, Singleton Park, Swansea, SA1 8PP, Wales, UK}
\affil[2]{{Department of Statistics}, {University of Warwick}, {{Coventry}, {CV4 7AL}, {England, UK}}}
\affil[3]{{Department of Pharmacology and Therapeutics}, 
{{University of Liverpool}, {Liverpool}, {L3 5TR}, {England, UK}}}
\affil[4]{{Division of Cancer and Genetics}, {Cardiff University}, {{Heath Park}, {Cardiff} {CF14 4XN, UK}}}
\affil[*]{Corresponding author. \href{email:j.m.rafferty@swansea.ac.uk}{j.m.rafferty@swansea.ac.uk}}

\maketitle
\markboth{BPR using VI for clustering MLTC}{BPR using VI for clustering MLTC}
\begin{abstract}
Multiple long-term conditions (MLTC) are increasingly observed in clinical practice globally. Clustering methods to group diseases into commonly co-occurring clusters have been of interest for further understanding of how MLTC group together and their associated impact on patient outcomes. However, such approaches require large, often population-scale datasets.
Bayesian Profile Regression (BPR) is a statistical model that combines a Dirichlet Process Mixture model with a hierarchical regression model, in order to form clusters of items conditional on covariates and an outcome of interest. We developed a BPR model using full-rank Stochastic Variational Inference (SVI) for application in large-scale data. We assessed it's performance using simulation studies comparing fits using the No-U-turn (NUTS) sampler and full-rank SVI. We then fit a BPR model to find clusters of MLTC in a population-scale data held in the Secure Anonymised Information Linkage (SAIL) databank.
We found results from full-rank SVI compared well with results from NUTS in a simulation study, and the improved fitting performance allowed for fitting models in population-scale datasets. There were 1,296,463 individuals in our electronic health record (EHR) cohort. The clustering model was conditioned on age at cohort entry, socioeconomic deprivation and sex with mortality as the outcome. We used the Elixhauser comorbidity index disease definitions, and found there were 33 disease clusters. 
We found that clusters featuring metastatic cancer and cardiovascular diseases, such as congestive heart failure, were most strongly associated with the probability of mortality. Our findings show that SVI can be a useful and accurate method for fitting Bayesian models, especially when the dataset size would make Monte Carlo methods prohibitively time consuming or impossible. 
\end{abstract}



\section{Introduction}

Multiple long-term conditions (MLTC) are increasingly experienced by people across the world, largely due to improvements in life expectancy and survival of acute health events \cite{johnston2019defining}. It has long been understood that MLTC can provide a difficult challenge to healthcare systems which are typically designed to address single conditions in isolation \cite{valderas2019quality}. Understanding how the development of long-term conditions changes the probability of accruing further diseases and modifies patient outcomes has proven to be an important but challenging prospect. There are many examples in the literature of efforts to find clusters of MLTC using various clustering methods including cluster analysis \cite{cornell2008multimorbidity}, latent class analysis \cite{hall2018multimorbidity, buja2018multimorbidity}, deep learning \cite{landi2020deep} and graph-based approaches \cite{held2016association, rafferty2021ranking,  burke2023representing, rafferty2023using}. These methods are typically applied in large scale, population level healthcare datasets in order to capture different combinations of disease. Examining all possible disease combinations is computationally very challenging, because the number of such combinations scales exponentially with the number of diseases considered.

Additional complexity is introduced when one considers identifying clusters of MLTC associated with an outcome, such as mortality. A simple 2-stage approach is commonly used \cite{siah2022multimorbidity} which identifies clusters using a clustering model in the first stage, and includes these clusters as an input to a regression model in the second stage. However, this approach would not be able to identify potentially rare clusters with an increased probability of outcome such as mortality or similar clusters resulting in different outcome probabilities. For example, there may be cases where we want to identify clusters of people with cancer, metastatic cancer and other associated conditions where people in one cluster experienced a high probability of death while people in the other cluster tended to recover. A unified, shared parameter modelling approach that identifies clusters (i.e. hierarchies) conditioned on a response is needed.

Bayesian profile regression (BPR) is a modelling framework that generates clusters based on a Dirichlet process mixture model which is also conditioned on an outcome using a generalised linear mixed-effects regression model (GLMM) \cite{molitor2010bayesian}. 
Previous work on BPR led to the development of the PReMiuM package, a library for R written in C++ that fits BPR models using Markov Chain Monte Carlo (MCMC) \cite{JSSv064i07} using a generalised slice sampler \cite{kalli2011slice} that allows for efficient Gibbs Sampling. MCMC approaches sample from the posterior distribution directly which has the advantage of enabling researchers to obtain diagnostic information about the fit of the model to ensure the samples are representative of the full posterior. However, when the dataset is large (as is the case when one wishes to analyse population-scale data) and / or the model is sufficiently complicated, MCMC can be intractable or prohibitively time consuming, even when using high-performance computers or computer clusters. The PReMiuM package does not currently support multi-threading and does not include support for analysing data in batches, meaning it cannot easily be used to analyse datasets with very large numbers of rows (representing individuals) or columns (diseases) due to processing time and / or memory requirements.  

Stochastic Variational Inference (SVI) is an alternative method which approximates the full posterior distribution with a tractable variational distribution or set of variational distributions, typically a mixture of Gaussian distributions that can be sampled from as if it were the true posterior \cite{blei2017variational}. SVI converts this inference problem into one of optimization. There has been much research on solving optimization problems in recent years in the context of machine learning research and as such SVI is often faster than MCMC approaches. SVI is particularly well suited to fitting models to large datasets where there is little chance of local minima, i.e. where there are not multiple candidate variational distributions which are a close match to the true posterior \cite{blei2017variational}. Care is needed to ensure the fitted variational distribution is an acceptable match for the true posterior, as with any such approximation methods as we are not sampling directly from the posterior \cite{wainwright2008graphical}.

The aim of this work was to develop a BPR model that could be applied to population-scale data to identify clusters of MLTC resulting in increased mortality adjusting for demographic variables. In section \ref{sec:methods}, we will describe in detail the construction and fitting of the model. In section \ref{sec:sim}, we will describe the data and present results from a simulation study and in section \ref{sec:sail}, we describe the data and present results from the model applied to a population-scale routine data problem. In section \ref{sec:discussion}, we will conclude with discussion and further work.

\section{Methods} \label{sec:methods}


\subsection{Model Likelihood}

The model likelihood, $p$, was composed of two components, that we refer to as the mixture component, $p_M$, and the response component, $p_R$. 
\begin{equation}
p = p_M . p_R
\end{equation}

\subsubsection{Mixture component $\left(p_M\right)$}

Let $z_i$  denote the cluster allocation for individual $i=1,\ldots,n$, where $z_i \in \left\{1,2,…n_k \right\}$ is distributed according to the mixture component likelihood $p_M$ with parameters $\pi_k$ for each cluster $k$. 
The mixture component of the model in this example is represented by a discrete Dirichlet process. 
\begin{equation}
    z_i \sim \mathrm{Discrete}\left( \left\{\pi_k\right\} \right)
\end{equation}
A Dirichlet Process Mixture Model (DPMM) \cite{ferguson1973bayesian, walker1999bayesian} is a model construction that groups observations into clusters, and can support any number of clusters $n_k$. In practice, the number of clusters is bounded by the number of observations in the dataset, since there cannot be more clusters than observations (assuming observations must belong to only one cluster). New observations have a probability of being added to an existing cluster, but also some non-zero probability of being added alone to a new cluster, and hence the number of clusters is variable. 

The prior probability of cluster membership, $\pi_k$ for each cluster $k$, is estimated using the stickbreak transformation \cite{ishwaran2001gibbs}. The stickbreak transformation takes a set of samples $v_k$ in the range $\left[0,1\right]$ and returns a set of probabilities $\pi_k$ for each cluster $k$, such that:
\begin{align}
    \pi_1 &= v_1 \\
    \pi_k &= v_k \prod_{\ell = 2}^{k - 1}\left(1 - v_\ell \right), \quad \mathrm{for}\, k \in \left\{2,3, \ldots, n_k-1 \right\} \nonumber\\
    \pi_{n_k} &= \prod_{\ell = 1}^{n_k - 1}\left(1 - v_\ell \right) \nonumber \\
    v_\ell &\sim \mathrm{Beta}\left(1, \alpha\right)  \nonumber\\
    \alpha &\sim \mathrm{Uniform}\left(0.3, 10\right) \nonumber
\end{align}
Let $x_{ij}=(x_{i1},\ldots,x_{ip})$ denote the $p$-dimensional covariate profile for individual $i$, and $\theta_k$ denote parameters for the distribution of covariates in cluster $k$, then the covariate model is defined as: 
\begin{equation}
    p_M\left(x_i |z_i=k \right)_i \sim f(x_i |\theta_k)
\end{equation}
Where $f$ is an appropriate probability distribution with parameters $\theta_k$, depending on the nature of $x_i$. 
In our example, we have a set of binary covariates, such that: 
\begin{align}
    p_M \left(x_{ij} |z_i=k \right)_{ij} &\sim \mathrm{Bernoulli}\left(\phi_j\right) \\
    \phi_j &\sim \mathrm{Uniform}\left(\epsilon, 1-\epsilon \right) \nonumber
\end{align}
The probability $\phi_j$ had an uninformative prior with $\epsilon$ set to a small positive value to prevent the exact values 0 and 1 from being sampled to improve numerical stability. 

\subsubsection{Response component $\left(p_R\right)$}
Let $y_i$  denote the response for individual $i=1,\ldots,n$, which is distributed according to the response model likelihood, $p_R$. In our example a logistic regression model was used to specify the response component using a binary outcome $y_i$ with cluster-specific intercepts $\beta_{0k}$ and global regression coefficients, $\beta_{a}$, for a set of response covariates $w_{ia}$, for each individual $i$ and response variable $a$, such that:
\begin{align}
p_R(y_i |z_i=k,w_{i})_i  &\sim \mathrm{Bernoulli}\left(Y_i \right) \\
\mathrm{logit}\left(Y_i \right) &= \beta_{0k} + \sum_a \beta_a w_{ia} \nonumber
\end{align}
The priors for all $\beta$ parameters of the response component were assigned a Student's t-distribution with 7 degrees of freedom, location parameter 0 and scale parameter of 2.5, in line with prior work \cite{molitor2010bayesian}.

\subsection{Model fitting}
For model fitting we used Numpyro \cite{bingham2019pyro, phan2019composable}, a Probabilistic Programming Language (PPL) for Python that uses JAX \cite{jax2018github} for tensor manipulations, and allows for several fitting methods, including the no-U-turn sampler (NUTS) and SVI. Discrete latent variables in the model were manually marginalised rather than sampled as part of the model fitting using both NUTS and SVI. Discrete latent variables were calculated from the posterior predictive distribution when needed, for example to perform posterior predictive checks.

The expected structure of the variational distribution (strictly, the class of variational distributions) approximating the full posterior was defined assuming that all parameters in the model (other than the discrete latent variables that were marginalised) were distributed according to a multivariate Normal distribution, which is often called the full-rank variational approximation \cite{kucukelbir2017automatic}. SVI was performed by maximising the Evidence Lower Bound (ELBO), equivalent to minimising the Kullback-Leibler (KL) divergence comparing the variational distribution and the full posterior distribution \cite{blei2017variational}. The means and covariance matrix $\Sigma_{ij}$ of the multivariate Normal distribution were tuned as part of the optimisation process. Probability parameters constrained to the interval $[0, 1]$, were transformed from the $\left(-\infty, \infty \right)$ range of the multivariate Normal variational distribution using a sigmoid transform. SVI fitting allows one to tune the optimisation with certain hyperparameters, the learning rate and the number of samples to compute the ELBO. The learning rate was set to 0.01 and the number of ELBO samples was set to 30. SVI, particularly approaches that utilise the mean-field approximation where the covariance matrix of the variational distribution is diagonal with zeros on the off-diagonal is known to underestimate parameter uncertainties, so to provide improved estimation of the full parameter uncertainties we used the full-rank variational approximation, which uses a (appropriately transformed) multivariate Normal as the variational distribution capturing the full covariance matrix and as such correlations between model parameters. 

SVI allows for models to be fit on a subset of the data at each step (a \textit{batch} of the data), rather than the full dataset. Since typically there are at least of order $10^4$ steps used in the SVI optimisation procedure, all data is used to fit the model even if a relatively small batch size is chosen. For our SVI workflow, at each step of the optimisation process a random batch data was sampled to be used for fitting. It was expected that reducing the batch size would reduce the time taken for the optimisation process to run, but would increase the variance of the computed ELBO. The number of steps needed for optimisation convergence was not expected to depend on the batch size. We found that a batch size at least 10\% of the dataset size was sufficient to ensure the optimisation process converged sufficiently. 

Dirichlet process mixture models exhibit the property of \textit{exchangeability} \cite{papaspiliopoulos2008retrospective, hastie2015sampling}. Briefly, in a DPMM, the label assignment for each cluster is permutation invariant, meaning the posterior is multimodal and leads to the potential problem of label-switching \cite{papaspiliopoulos2008note}. Label-switching can occur separately across chains, which can be resolved by relabelling the clusters after sampling as a post-processing step, or within chains. Label-switching within chains has been addressed previously using custom sampling algorithms \cite{papaspiliopoulos2008retrospective, hastie2015sampling} or by imposing an order on parameters. Label-switching is more likely to be observed when dealing with continuous mixtures, where the multiple distributions overlap. In the discrete case the most commonly observed label switching behaviour is across chains. As such, the re-labelling post-processing step was implemented, and trace plots were carefully checked to ensure no within-chain label-switching occurred. For SVI fits, since the variational distribution assumed unimodality it was not possible for the results from SVI to feature multimodality.

%
%

Analysis code was written in Python version 3.12.6 using numpyro version 0.15.0 \cite{phan2019composable}, JAX version 0.4.28 \cite{jax2018github}, arviz version 0.18.0 \cite{Kumar2019}, numpy version 1.26.4 \cite{harris2020array} and pandas version 2.2.2 \cite{the_pandas_development_team_2024_10957263}. Analysis code written for this work is available at: \url{https://github.com/jim-rafferty/BPR_MLTC_clustering}.

\section{Simulation Study} \label{sec:sim}

\subsection{Data}

A simulation study was performed in accordance with guidance set out previously \cite{burton2006design}. A preliminary simulation study was performed to determine the sample size needed utilising the criteria set out in section 2.6 of \cite{burton2006design}, which was found to be $N=8000$ observations and $M=800$ simulations. We performed simulations comparing NUTS and SVI with $N=8000$ observations, and simulations fit using only SVI with $N=100,000$ observations. Data was simulated with $D_M = 12$ binary mixture variables, $D_R = 6$ response variables, one of which was continuous and the remainder binary, and $n_k=5$ clusters with probabilities $\left\{0.833, 0.166, 0.25, 0.2083, 0.2916\right\}$, chosen to be adequately distinguishable in the results.

Data for the mixture component of the model $x_{ij}$ were drawn from 12 independent Bernoulli distributions for each simulation. The Bernoulli probabilities $\phi_{k j}$ were chosen such that for each cluster two variables were chosen to have relatively higher probability than the others. Probabilities for the other variables in each cluster were chosen to be small, but constrained to be larger than 0.01.
This simulated individuals with a small baseline probability of having any disease, while a higher probability of having a specific disease implied a higher probability of having another specific disease as well. For simplicity, each cluster had a high probability of disease that was in common with one other cluster. Finally, a $N \times D_d$ array of samples were drawn from $D_d$ independent Bernoulli distributions:

\begin{equation}
x_{i} \sim \prod_j \mathrm{Bernoulli}\left(\phi_{k j} \right)
\end{equation}

For the response component of the data $w_{ia}$, one variable was generated from a Normal distribution $\mathcal{N}\left(45, 100\right)$, to represent an age variable. The remaining variables were drawn from Bernoulli distributions to represent sex and deprivation quintile variables. The gradients $\beta$ were chosen to approximate a realistic scenario where increasing age, male sex and increased deprivation increased the probability of the outcome, which in this case was representative of mortality.
Model intercepts $\beta_{0k}$ were selected for each cluster $k$ to ensure a realistic number of observations of the outcome would occur in the overall dataset, and would differ between clusters. The outcome was generated from a Bernoulli distribution such that

\begin{equation}
y_{i} \sim  \mathrm{Bernoulli}\left[\mathrm{logistic} \left(\beta_{0 k} + \sum_{a = 1}^{D_R}\beta_{a} w_{ia} \right) \right]
\end{equation}
where the logistic function is the inverse of the logit function. The order of rows in the data table were randomised and continuous variables were centered prior to fitting the model. The mean bias and coverage was calculated:
\begin{equation}
    b = \frac{1}{M}\sum_{i = 1}^{M}{\left(\theta_i - \hat{\theta}\right)}
\end{equation}
for $\theta = \beta_r$ and $\theta = \phi_{k j}$ where $\hat{\theta}$ denotes the parameter estimate. For the mixture component $\phi$ parameters which are interpreted as probabilities, the bias in the log-odds was used such that:
\begin{equation}
    \log \left( \mathrm{odds} \right) = \log\left(\frac{\phi}{1 - \phi}\right)
\end{equation}
The coverage was calculated as the proportion of times the true parameter fell within the 95\% credible interval, computed using the Highest Density Interval (HDI) method.

For the main simulation study, results from SVI and NUTS were compared. We performed additional simulation studies to explore the regime of validity of the SVI method. We examined the effect of batch size on goodness of fit with the simulation parameters described above. Furthermore, a simulation study with $N=100,000$ was performed to ensure results were replicated at large scale. It was not possible to perform the simulation study with $N = 100,000$ using NUTS, as it would be time prohibitive and/or computationally intractable. Simulation studies were performed on a workstation computer with 16 hardware threads. 

\subsection{Results} \label{sec:results}

\begin{figure}
    \centering
    \includegraphics[width=0.49\linewidth]{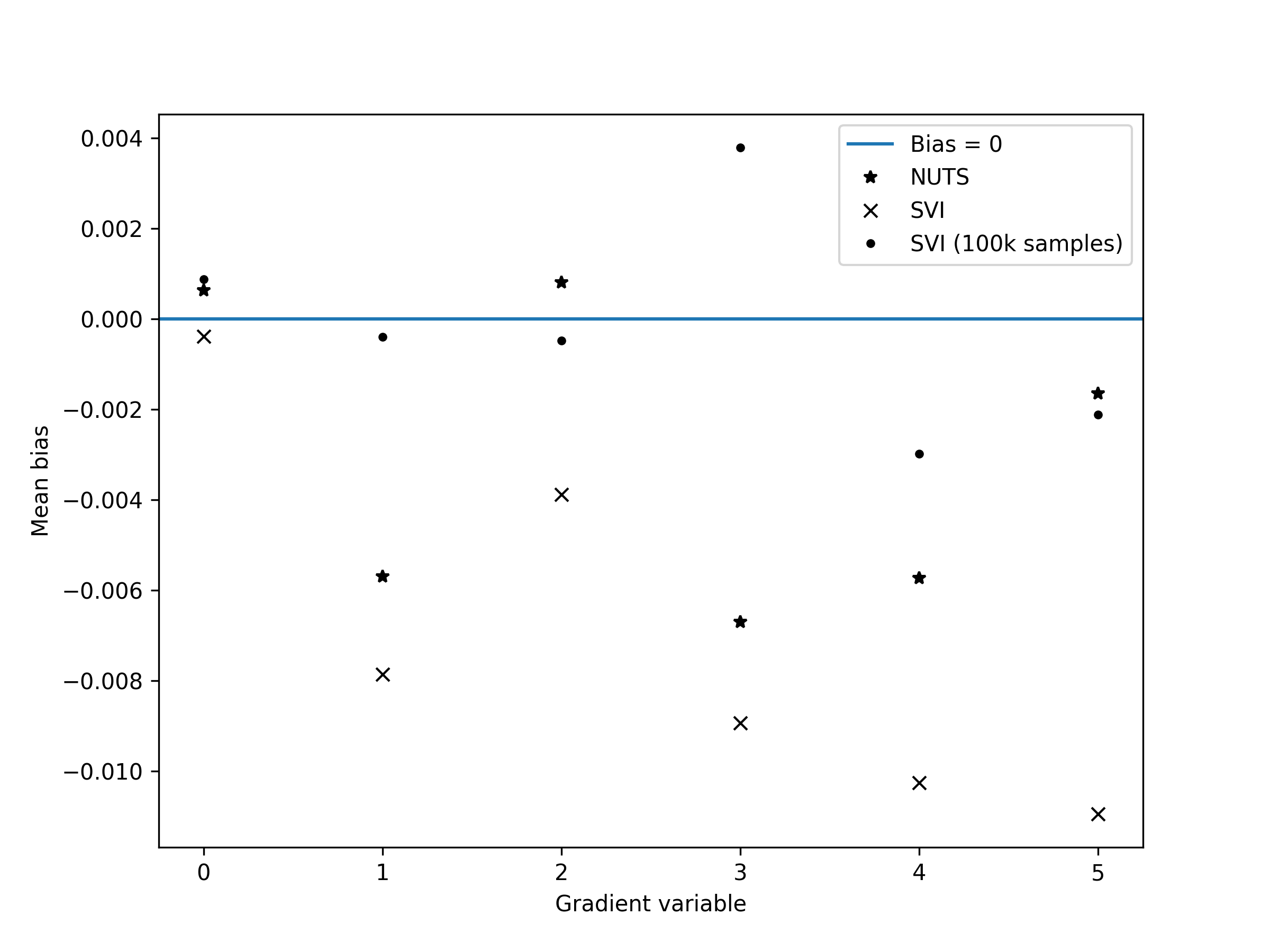}
    \includegraphics[width=0.49\linewidth]{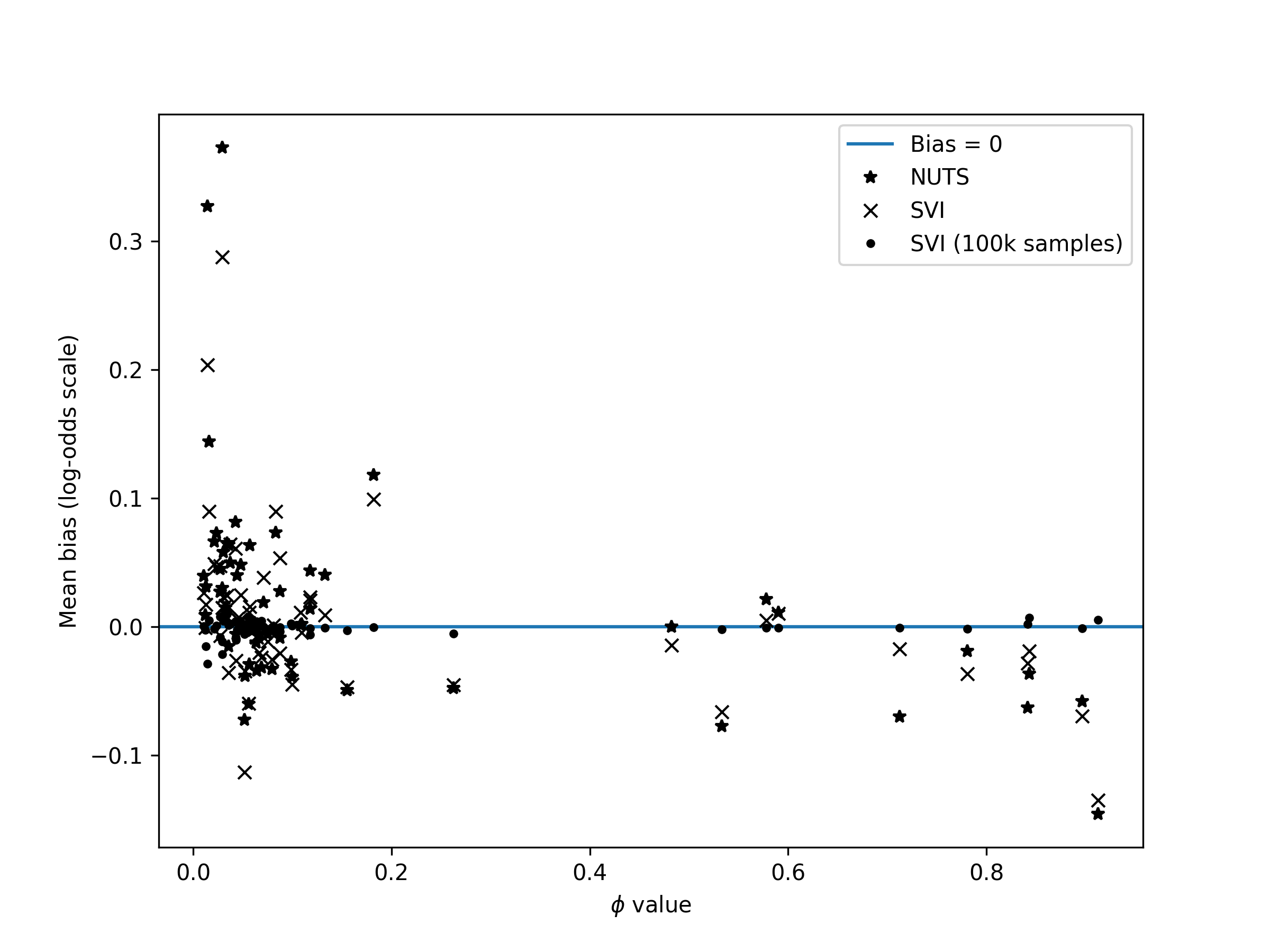}\\
    \includegraphics[width=0.49\linewidth]{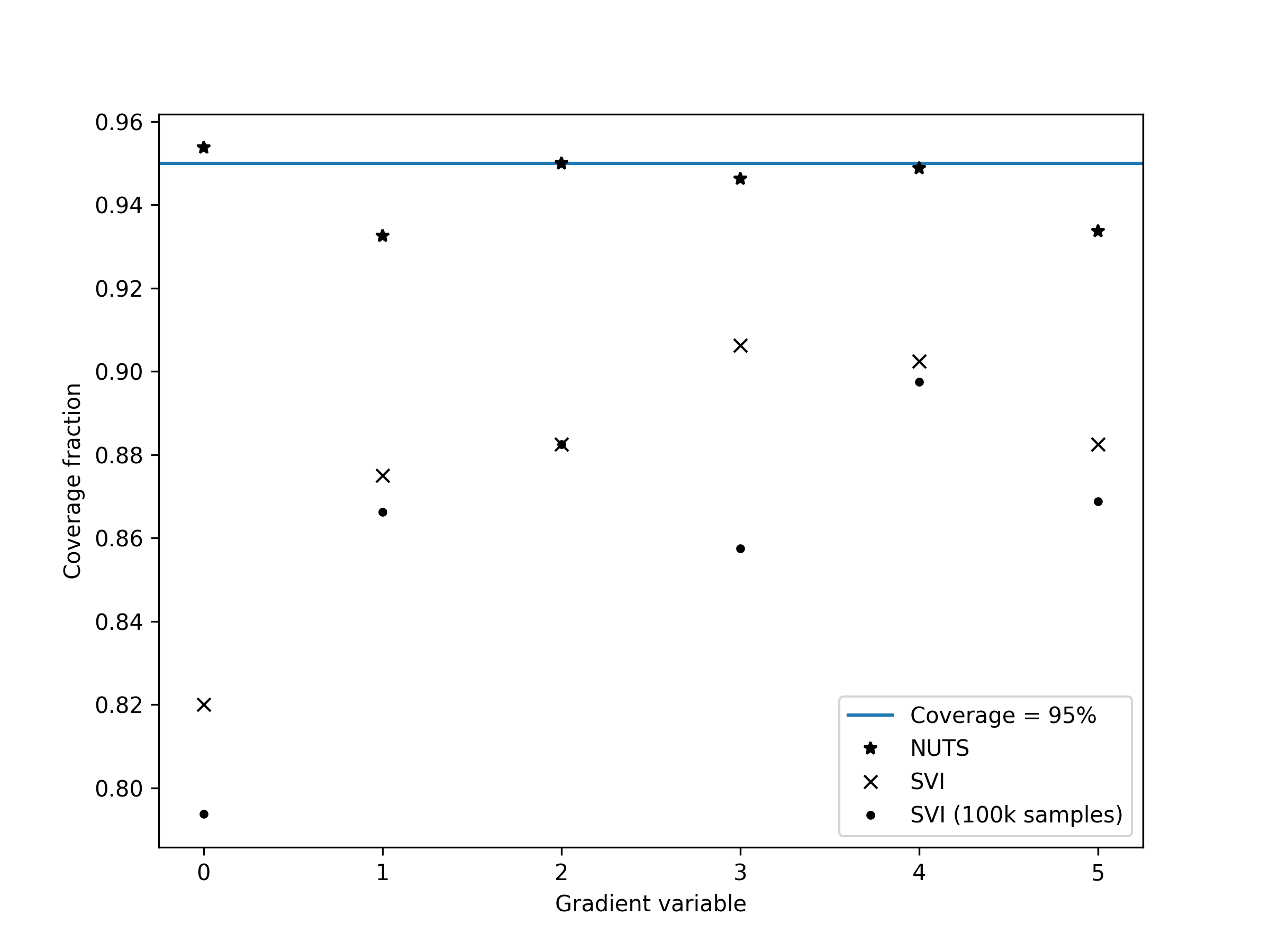}
    \includegraphics[width=0.49\linewidth]{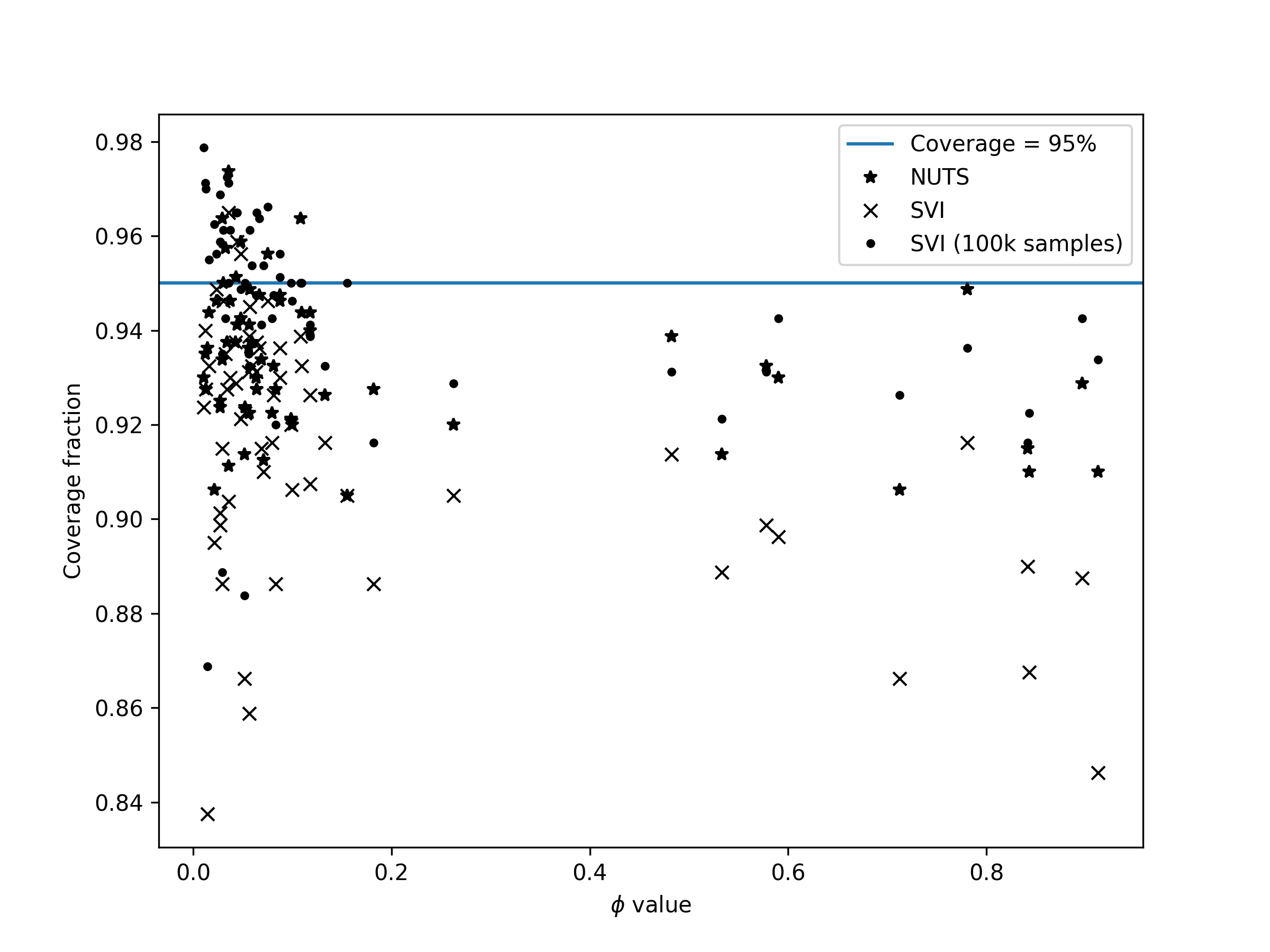}
    \caption{Parameter bias and 95\% coverage computed by simulation. Top left: Biases of the estimated gradients in the response component. Top right: Biases of estimated probabilities in the mixture component on the log-odds scale, plotted against the value of the $\phi$ parameter. Bottom left: Coverage estimates of gradient parameters in the response component. Bottom right: Coverage estimates in the mixture component, plotted against the value of the $\phi$ parameter.}
    \label{fig:sim_results}
\end{figure}

Results from the three sets of simulation studies are shown in Figure \ref{fig:sim_results}. Biases in the response component gradient and mixture component $\phi$ parameters are shown in Figure \ref{fig:sim_results} (top row). We observed that smaller $\phi$ parameters were more challenging to estimate, with biases for $\phi \gtrsim 0$ larger than for larger $\phi$ values for all fit methods. Results for coverage estimates from the simulation study are presented in Figure \ref{fig:sim_results} (bottom row). The coverage for the response component gradient parameters was slightly smaller than expected for SVI fitting with coverage consistently above 80\%, which is expected given the approximate nature of the method. Coverage tended to be lower for response component gradient estimates in datasets of a larger sample size while for the mixture component $\phi$ parameters coverage was improved for larger datasets. Overall, the mean bias in the parameters was small in magnitude and close to 0. 

\begin{figure}
    \centering
    \includegraphics[width=0.49\linewidth]{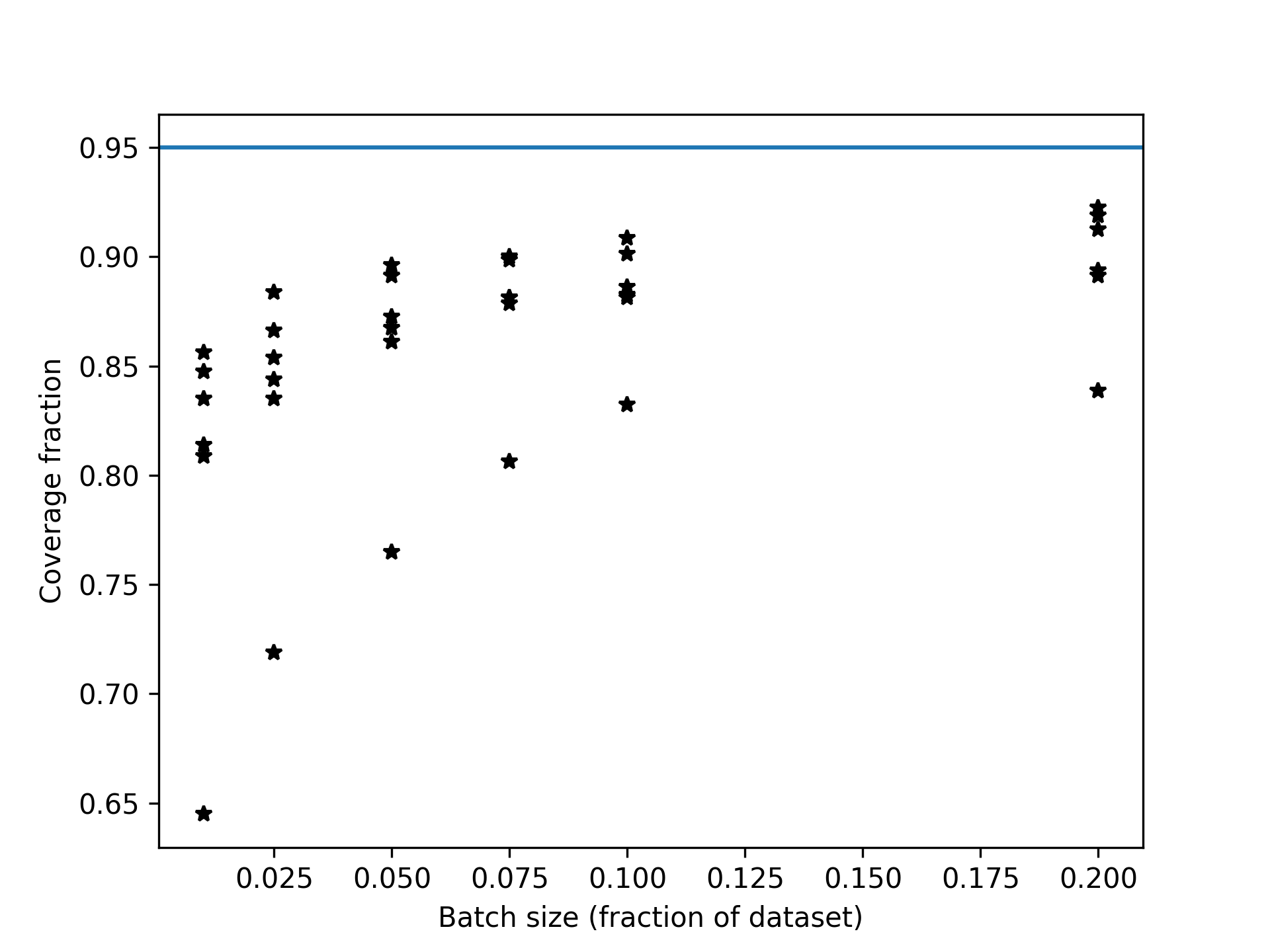}
    \includegraphics[width=0.49\linewidth]{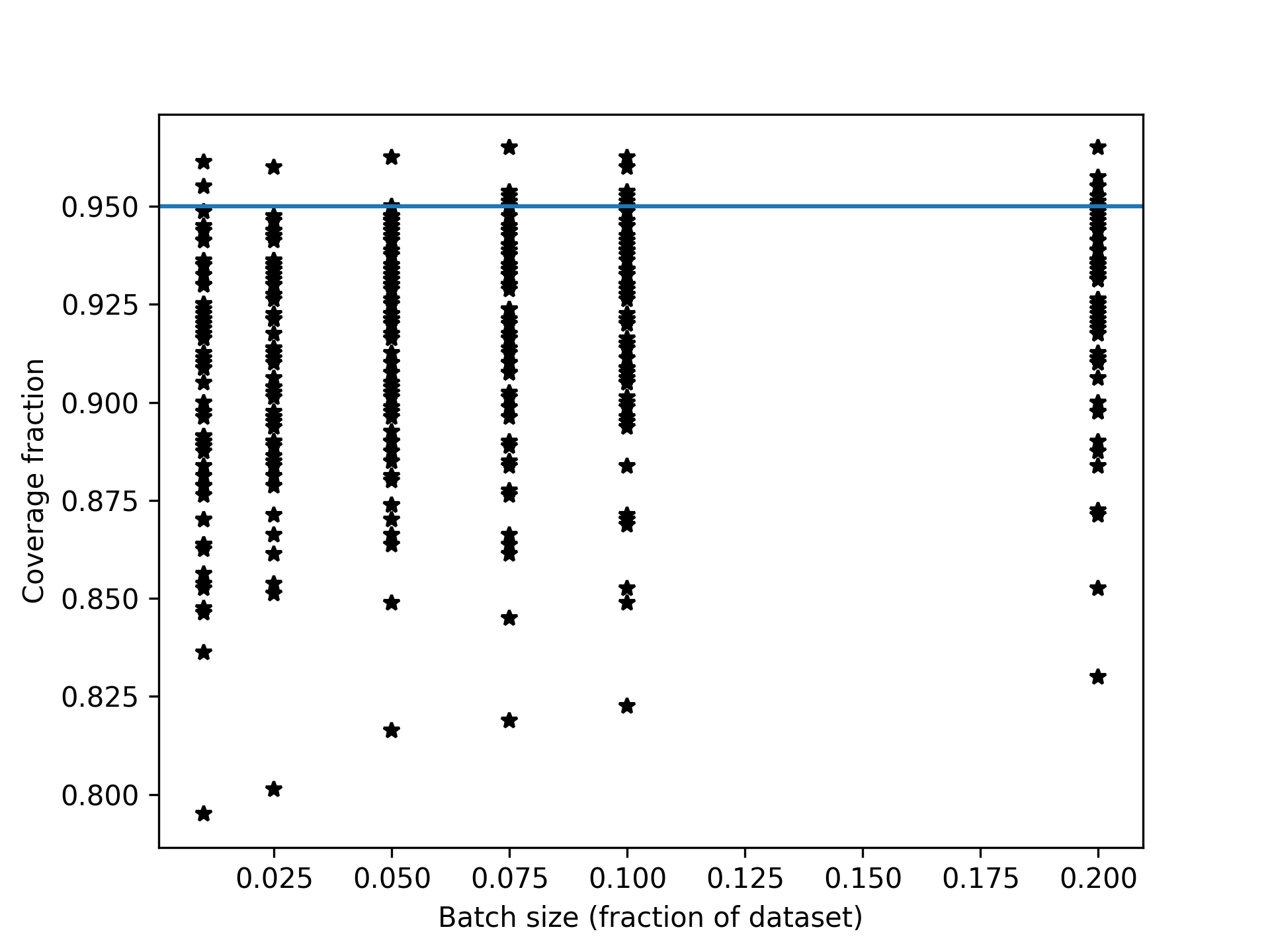}
    \caption{Parameter 95\% coverage estimates by simulation as a function of batch size in SVI fits. Left: Coverage estimates for the gradient parameters in the response component. Right: Coverage estimates for the $\phi$ parameters in the mixture model. Horizontal lines show coverage = 95\%}
    \label{fig:sim_results_multibatch}
\end{figure}

Results of simulations varying the batch size showed that the bias did not depend on the batch size. Coverage did depend on the batch size and was broadly optimised by using at least 10\% of the dataset in each batch. Simulation results showing how coverage varies with batch size are shown in Figure \ref{fig:sim_results_multibatch}.

\section{Application to Routine Healthcare Data} \label{sec:sail}

\subsection{Data}
We used Electronic Health Record (EHR) and administrative data held in the Secure Anonymised Information Linkage (SAIL) Databank, a privacy protecting Trusted Research Environment \cite{ford2009sail, lyons2009sail}. The SAIL Databank contains all secondary care records and approximately 85\% of primary care records for individuals receiving care in Wales, UK \cite{thayer2020measuring}. Approval for the use of anonymised data in this study, provisioned within the SAIL Databank, was granted by an independent Information Governance Review Panel (IGRP) under project 1744. We utilised the Wales Multimorbidity e-cohort (WMC) which includes all people aged 18 years or older alive and living in Wales on $1^\mathrm{st}$ January 2000 \cite{lyons2021protocol}. We used disease diagnoses as the variables for fitting the mixture component of the model, with diseases defined using the Elixhauser comorbidity index \cite{elixhauser1998comorbidity, metcalfe2019coding}. A clearance period of 5 years was applied to ensure that only new diagnoses were considered. The response component of the model utilised age at cohort entry (with cohort mean age subtracted to improve model fitting performance), sex and Welsh Index of Multiple deprivation (WIMD) quintile, a geographical measure of multiple deprivation provided by the Welsh government \cite{WIMD_technical_report}, with the third deprivation quintile chosen as the reference. The outcome was mortality, as provided by the Office for National Statistics. As a subgroup analysis we also fitted the model to the data stratified by age at cohort entry in 20 year strata. 

\subsection{Results}
There were 1,296,463 individuals included in the analysis. The mean age was 42.51 (SD: 15.95) years. 48.7\% of the cohort were male. There was a slight imbalance in proportions of the population in the five WIMD quintiles (17.99\% of the population was in the most deprived quintile, while 21.31\% was in the least deprived quintile). See Table \ref{tab:summary} for a summary of the cohort profile. The most frequently observed disease was hypertension, affecting 11.48\% of the cohort, followed by cancer and Chronic Obstructive Pulmonary Disease (COPD) affecting 5.98\% and 5.06\% of the cohort respectively. The least commonly observed diseases were psychosis (0.27\%), coagulopathy (0.24\%) and blood loss anaemia (0.08\%). See Table \ref{tab:summary_diseases} for information on prevalence of the diseases present in the cohort. 

\begin{table}[h!]
    \centering
    \begin{tabular}{ll}
    \toprule
  Variable &  N(\%) / Mean(SD)\\
  \midrule
  Population N & 1296463 \\
  Male Sex N(\%) & 631453 (48.70\%)\\
  Age Mean (SD) (years) & 42.51 (15.95) \\
  WIMD quintile = 1 (most deprived) N(\%) & 233243 (17.99\%)\\ 
  WIMD quintile = 2 N(\%) & 253161 (19.53\%)\\ 
  WIMD quintile = 3 N(\%) & 264308 (20.39\%)\\ 
  WIMD quintile = 4 N(\%) & 269474 (20.79\%)\\ 
  WIMD quintile = 5 (least deprived) N(\%) &276277 (21.31\%)\\ 
  Died N(\%) & 160709 (12.40\%) \\
  Follow up Mean (SD) (years / person) & 16.16 (6.49) \\ 
  Total Follow up (years) & 20949059\\
  \bottomrule
    \end{tabular}
    \caption{A summary of the demographic and mortality variables used in the response component of the model, as well as follow-up information. }
    \label{tab:summary}
\end{table}

\begin{table}[h!]
    \centering
    \begin{tabular}{ll}
    \toprule
      Disease  & Count(\%) \\
      \midrule 
            Hypertension & 148781 (11.48\%) \\
            Cancer & 77483 (5.98\%) \\
            Chronic Obstructive Pulmonary Disease (COPD) & 65566 (5.06\%) \\
            Arrhythmia & 61173 (4.72\%) \\
            Depression & 56840 (4.38\%) \\
            Diabetes & 54368 (4.19\%) \\
            Other neurological disorder (OND) & 41392 (3.19\%) \\
            Fluid \& electrolyte disorders & 35925 (2.77\%) \\
            Obesity & 33386 (2.58\%) \\
            Metastatic cancer & 31849 (2.46\%) \\
            Renal failure & 31649 (2.44\%) \\
            Congestive heart failure & 28378 (2.19\%) \\
            Deficiency anaemia & 27814 (2.15\%) \\
            Hypothyroidism & 24101 (1.86\%) \\
            Weight loss & 22300 (1.72\%) \\
            Valvular disease & 21940 (1.69\%) \\
            Peripheral vascular disease & 18270 (1.41\%) \\
            Liver disease & 12936 (1.00\%) \\
            Diabetes with complication & 11888 (0.92\%) \\
            Pulmonary circulation disorders & 11246 (0.87\%) \\
            Rheumatoid arthritis & 11307 (0.87\%) \\
            Peptic ulcer & 9017 (0.70\%) \\
            Paralysis & 6919 (0.53\%) \\
            Lymphoma & 4318 (0.33\%) \\
            Hypertension with complication & 3717 (0.29\%) \\
            Drug abuse & 3779 (0.29\%) \\
            Psychosis & 3470 (0.27\%) \\
            Coagulopathy & 3080 (0.24\%) \\
            Blood loss anaemia & 1043 (0.08\%) \\
            \bottomrule
    \end{tabular}
    \caption{Information on the prevalence of the diseases in the cohort, used in the mixture component of the model. Note that disease phenotypes that can be with or without complication (hypertension, diabetes) are mutually exclusive categories - the same individual cannot have both diabetes and diabetes with complication for example.}
    \label{tab:summary_diseases}
\end{table}

\begin{figure}[h!]
    \centering
    \includegraphics[width=0.8\columnwidth]{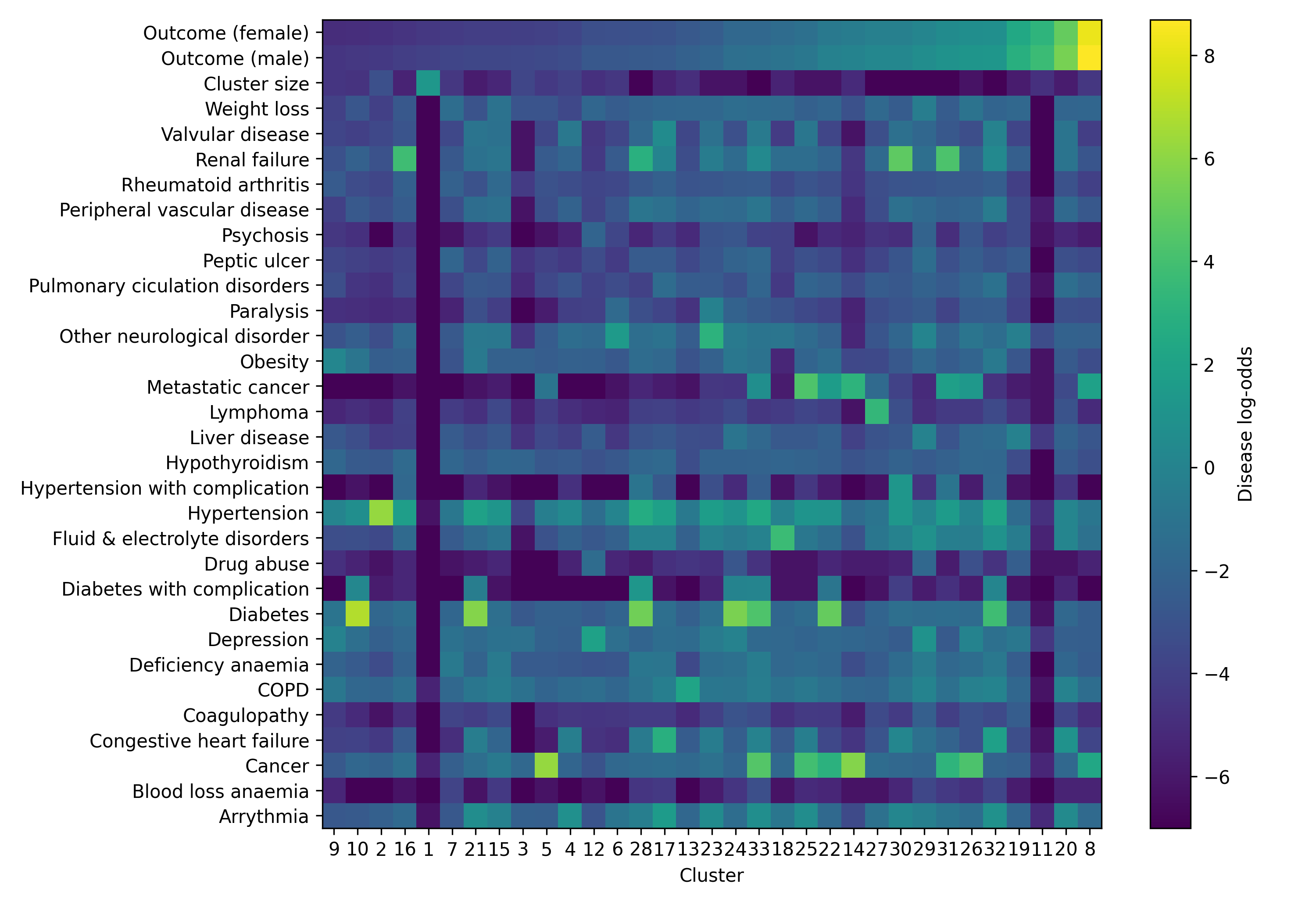}
    \caption{A heatmap showing point estimates for the log-odds of disease and the outcome for each cluster. The outcome is calculated for women and men at the mean age of 42.5 years and living in the third deprivation quintile. A plot with the colours on a linear scale is in Supplemental Section \ref{sec:lin_scale}.}
    \label{fig:SAIL_probs}
\end{figure}

We fitted the model with a maximum of 50 clusters and found there were 33 non-empty clusters. The global gradients showed patterns one would expect, in that increased age (odds ratio [OR]: 1.154, 95\% credible interval [CrI] (1.153, 1.156)) and the most deprived individuals had increased odds of mortality (OR: 1.40, 94\% CrI (1.38, 1.44)). Women compared to men (OR: 0.62 94\% CrI (0.61, 0.62) and the individuals in the least deprived quintile (OR: 0.71 94\% CrI (0.70, 0.72)) had reduced odds of mortality. Disease, mortality and cluster probabilities are shown in Figure \ref{fig:SAIL_probs}. Coefficients and intercepts are shown in Table \ref{tab:coefficients}, cluster probabilities are shown in Table \ref{tab:cluster_proba}.

\begin{table}[h!]
    \centering
    \small
    \begin{tabular}{lll}
    \toprule 
    Variable &  Coefficient estimate (95\% CrI) & Odds Ratio (95\% CrI) \\
    \midrule				
Age               & 0.144  (0.143, 0.145)    & 1.154 (1.153, 1.156) \\
Female Sex        & -0.483 (-0.495, -0.471) & 0.62 (0.61, 0.62) \\
WIMD Quintile = 1 & 0.340  (0.320 , 0.362) & 1.40 (1.38, 1.44) \\
WIMD Quintile = 2 & 0.194  (0.175 , 0.213) & 1.21 (1.19, 1.24) \\
WIMD Quintile = 4 & -0.136  (-0.154, -0.116) & 0.87 (0.86, 0.89) \\
WIMD Quintile = 5 & -0.344 (-0.362, -0.322) & 0.71 (0.70, 0.72) \\
\midrule
Cluster &  Intercept estimate (95\% CrI) & Outcome probability (95\% CrI)\\
\midrule 
Cluster 1 & -3.956 (-3.975, -3.938) & 0.783 (0.782, 0.784)  \\
Cluster 2 & -4.373 (-4.418, -4.325) & 0.041 (0.040, 0.042)  \\
Cluster 3 & -3.630 (-3.744, -3.523) & 0.024 (0.023, 0.026)  \\
Cluster 4 & -3.329 (-3.388, -3.269) & 0.018 (0.017, 0.018)  \\
Cluster 5 & -3.467 (-3.567, -3.375) & 0.012 (0.011, 0.013)  \\
Cluster 6 & -3.790 (-3.895, -3.684) & 0.011 (0.010, 0.012)  \\
Cluster 7 & -2.717 (-2.817, -2.627) & 0.011 (0.010, 0.011)  \\
Cluster 8 & 8.695 (6.429, 10.963) & 0.011 (0.010, 0.011)  \\
Cluster 9 & -4.559 (-4.785, -4.320) & 0.010 (0.009, 0.012)  \\
Cluster 10 & -4.477 (-4.605, -4.314)  & 0.009 (0.009, 0.010) \\
Cluster 11 & -2.735 (-2.911, -2.563)  & 0.008 (0.007, 0.009) \\
Cluster 12 & 3.685 (3.497, 3.905)  & 0.008 (0.007, 0.008) \\
Cluster 13 & -2.183 (-2.293, -2.049)  & 0.007 (0.007, 0.008) \\
Cluster 14 & 0.031 (-0.153, 0.210)  & 0.006 (0.006, 0.006) \\
Cluster 15 & -3.683 (-3.820, -3.535)  & 0.005 (0.005, 0.006) \\
Cluster 16 & -1.006 (-1.258, -0.728)  & 0.004 (0.004, 0.004) \\
Cluster 17 & -2.532 (-2.668, -2.414)  & 0.004 (0.003, 0.004) \\
Cluster 18 & -4.155 (-4.285, -4.029)  & 0.004 (0.004, 0.004) \\
Cluster 19 & 2.882 (2.643, 3.130)  & 0.003 (0.003, 0.003) \\
Cluster 20 & -3.691 (-3.867, -3.536)  & 0.003 (0.003, 0.004) \\
Cluster 21 & 5.457 (3.873, 7.146)  & 0.003 (0.003, 0.004) \\
Cluster 22 & -1.925 (-2.405, -1.476)  & 0.002 (0.001, 0.002) \\
Cluster 23 & 1.169 (0.760, 1.565)  & 0.002 (0.001, 0.002) \\
Cluster 24 & -0.789 (-0.939, -0.632)  & 0.002 (0.002, 0.002) \\
Cluster 25 & -1.309 (-1.646, -0.931)  & 0.002 (0.001, 0.002) \\
Cluster 26 & -0.165 (-0.350, 0.013)  & 0.002 (0.002, 0.002) \\
Cluster 27 & -2.633 (-2.921, -2.326)  & 0.001 (0.001, 0.001) \\
Cluster 28 & 0.233 (-0.019, 0.504)  & 0.001 (0.001, 0.001) \\
Cluster 29 & 1.017 (0.433, 1.553)  & 0.001 (0.001, 0.001) \\
Cluster 30 & 0.645 (0.173, 1.050)  & 0.001 (0.001, 0.001) \\
Cluster 31 & 0.253 (-0.174, 0.665)  & 0.001 (0.001, 0.002) \\
Cluster 32 & 1.217 (0.430, 1.849)  & 0.001 (0.001, 0.001) \\
Cluster 33 & -1.281 (-1.686, -0.865) & 0.000 (0.000, 0.000) \\
\bottomrule
\end{tabular}
    \caption{Coefficients and intercepts from the BPR model fit to routine healthcare data. The outcome probability for each cluster is the probability of the outcome for men at the mean cohort age (42.51 years) living in the third deprivation quintile. CrI - Credible interval (Highest Density Interval, HDI), WIMD quintile = 1 is the most deprived, WIMD quintile = 5 is the least deprived.}
    \label{tab:coefficients}
\end{table}
The cluster with the highest probability of cluster membership (cluster 1, 78.3\%) included people who were unlikely to have any chronic diseases as defined by the Elixhauser comorbidity index and had a small probability of mortality (1.9\% for males and 1.2\% for females of average age in the middle deprivation quintile). This cluster is representative of the majority of the population who are relatively disease free with a small probability of mortality. A table of all cluster membership probabilities is in Supplemental Section \ref{sec:cluster_proba}.

Individuals in the cluster with the third highest probability of mortality (cluster 11, representing 0.8\% of the cohort) had small probabilities of all diseases, in contrast to almost all of the other clusters. The diseases with the highest probability of affecting individuals in this cluster were other neurological disorders (OND) (3.4\%), which includes many neurological diseases such as cerebrovascular disease, prior stroke, Parkinson's disease, dementia and epilepsy (see \cite{metcalfe2019coding} for a full list of diseases included in this phenotype) followed by liver disease (1.3\%), depression (1.1\%) and hypertension (0.8\%). 

We note the existence of several clusters where there was a high probability of both cancer and metastatic cancer (clusters 8, 14, 22, 25, 26, 31 and 33), but only one cluster where there was a high probability of cancer but a lower probability of metastatic cancer (cluster 5). Cluster 8, representing 1.1\% of the cohort, in particular is interesting because it has relatively high probability of cancer and metastatic cancer (91.3\% and 87.4\% respectively) but relatively low probabilities of other diseases (the most probable being hypertension, 30.1\%, fluid and electrolyte disorders [FED], 22.9\% and chronic obstructive pulmonary disease, 19.3\%) coupled with the highest outcome probability of any cluster (99.98\% for males and 99.97\% for females at the mean cohort age). Cluster 25, representing 0.2\% of the cohort, had a much smaller outcome probability (31.2\% for males and 21.9\% for females at the mean cohort age) experienced cancer (98.1\%) and metatstatic cancer (98.7\%) as well as other diseases with high probability (hypertension, 75.6\%,  arrythmia, 64.7\%, and congestive heart failure [CHF], 41.5\%)

The highest probability of any disease in any cluster was the probability of diabetes in cluster 10 (99.9\%). This cluster represented 0.9\% of the cohort and had a probability of mortality of 1.1\% for males and 0.7\% for females at the mean cohort age. Other diseases occurring with high probability in this cluster were hypertension (66.2\%), diabetes with complication (56.4\%), obesity (29.0\%) and depression (20.4\%). Diabetes was present with high probability in several other identified clusters (21, 22, 24, 28, 32 and 33). Of these clusters, cluster 32 has the highest probability of mortality (77.2\% for males and 67.6\% for females at the mean cohort age), and high probabilities of hypertension (89.6\%), CHF (86.5\%), FED (73.2\%), arrythmia (70.2\%), renal disease (60.2\%), diabetes with complication (54.1\%), COPD (51.0\%), valuvular disease (48.3\%) and peripheral vascular disease (36.9\%) compared to other clusters.

There were three clusters with a relatively high probability of depression (cluster 12, representing 0.8\% of the cohort, where 87.9\% had depression, cluster 29, 0.1\% of the cohort, 72.5\% of which had depression and cluster 26, 0.2\% of the cohort, 50.4\% of which had depression). In cluster 12, the next most likely disease other than depression was COPD (19.5\%). In cluster 29 other common diseases were FED (70.3\%), hypertension (52.8\%), OND (51.5\%), COPD (49.9\%) and liver disease (47.4\%). In cluster 26 depression was no longer the most probable disease with cancer (98.6\%) and metastatic cancer (79.9\%) being more probable, with other probable diseases including hypertension (49.7\%) and COPD (44.7\%). The outcome probabilities for cluster 12, 29 and 26 for men at the mean cohort age was 6.1\%, 65.6\% and 76.1\% respectively, while for women at the mean cohort age was 3.8\%, 54.0\% and 66.5\% respectively. 

Results from a subgroup analysis of the cohort stratified by age are presented in Supplemental Section \ref{sec:age_strat}.

\section{Discussion} \label{sec:discussion}

Our findings have demonstrated that SVI can be used to fit BPR models to large-scale data. NUTS is computationally expensive, and for population-scale datasets such as the one presented in Section \ref{sec:sail}, it is not possible to use NUTS to sample model parameters. Our results show that for this example, SVI is an efficient alternative approximation to directly sampling the posterior distribution using MCMC methods.

Our simulation study has shown that directly sampling the posterior using NUTS produces small biases. The biases produced with SVI fitting are not generally larger but are more variable than those observed for NUTS. Biases for both methods were centered on zero, indicating the SVI fitting process does not introduce systematic biases into estimated parameters. The biases in the $\phi$ parameters of the mixture component when they are close to zero or one do show a small systematic bias with both NUTS and SVI fitting methods with smaller sample sizes, suggesting that the $\phi$ parameter is more difficult to estimate at the extremes. In the case when the true $\phi$ parameters were small the coverage was closer to 95\%, indicating the model correctly finds larger uncertainties for $\phi \gtrapprox 0$. In the case of $\phi$ parameters close to one, larger sample sizes accessible when fitting using SVI improved the bias and coverage estimates of the $\phi$ parameters dramatically and compared favourably to fits of smaller datasets using both fit methods.

To our knowledge, our study features the largest cohort of people to be studied using a BPR model. BPR models are very useful for understanding complex multiple long-term conditions as they form clusters based on disease coincidence but are also conditioned on variables that one may want to adjust for such as deprivation, sex and age, and on an outcome variable such as mortality. This additional model structure allows for the identification of clusters with similar diseases but different outcomes which, as observed in our study, is an important distinction. Furthermore, we are not aware of any previous study that has used variational inference to fit such a model. Such methods are crucial in enabling these models to be used to explore large, often population-scale, datasets that are a feature of modern epidemiological research. This research opens a collection of applications of BPR to population-scale datasets, where it was previously unfeasible to fit such models, such as identifying clusters of medications resulting in medications related harm.

Since our results feature clusters conditioned on mortality, the clusters that we have identified are not necessarily the same as those identified in previous studies.  Previous studies have identified a musculoskeletal disease cluster \cite{freund2012patterns, john2003patterns}. The only musculoskeletal disease in our chosen disease definitions is rheumatoid arthritis, which was relatively rare in our cohort and which typically causes excess mortality via mediating conditions such as cardiovascular disease \cite{sokka2008mortality}.

We found several clusters of interest. The largest cluster, cluster 1, represented the majority of the cohort with low recorded comorbidity and low mortality risk. Cluster 11 featured low probability of disease but relatively high probability of mortality. One plausible explanation is that it captures deaths due to trauma, accidents, suicide, or sudden acute medical events in individuals who otherwise have relatively low recorded long-term morbidity, which could be due to a genuinely low burden of chronic disease or due to individuals with low engagement with healthcare services. It is also possible that individuals in this cluster have diseases which are not included in the Elixhauser comorbidity index but are important causes of mortality such as coronary artery disease. This would represent a limitation of our use of the Elixhauser comorbidity index for disease definitions in our study. Finally, variable capture of conditions across care settings may lead to under-ascertainment of chronic disease and thus mask the true clinical complexity of this group.

There are several clusters featuring cancer. Cancer is a highly heterogeneous group of diseases where mortality is multifactorial, depending on cancer type, stage, biology, comorbidities, and treatment-related toxicity. A particularly important clinical distinction is between early-stage and potentially curable disease and metastatic disease which is generally incurable. Several clinical trajectories are possible; patients may truly have early-stage disease; they may be initially thought to have early-stage disease but are found to have metastases during staging; or they may have previously undetected metastatic disease.
The separation between Cluster 5 (high cancer probability, low metastatic cancer probability) and Cluster 8 (high cancer and high metastatic cancer probability) highlights the distinction between early and late disease, as defined by the Elixhauser comorbidity index's cancer and metastatic cancer phenotypes. Cluster 5 may predominantly represent early-stage or more indolent malignancies (e.g. localised prostate or breast cancer), or cancers with better treatment options, consistent with the observed lower mortality. Conversely, Cluster 8, with extremely high mortality, may represent advanced-stage malignancies with limited effective treatment options (e.g. pancreatic cancer or glioblastoma), or patients presenting late with widespread disease. Other metastatic cancer–containing clusters (such as Clusters 25 and 33) may illustrate the modifying effect of comorbidity on mortality risk. Cardiovascular disease (e.g. congestive heart failure) may restrict cancer treatment options and also affect non-cancer mortality, while shared risk factors such as smoking and obesity may drive the co-occurrence of cancer and cardiometabolic disease. Cluster 25 appears consistent with a ``cancer–cardiovascular disease'' phenotype, while Cluster 33 may reflect a more complex phenotype involving cancer, metastatic disease, and diabetes, leading to worse overall outcomes. Overall, the BPR model has enabled a more granular assessment of cancer clusters by conditioning on mortality as an outcome, which would otherwise not be identifiable from a 2-stage approach.

The model finds two phenotypes involving diabetes, which due to our use of the Elixhauser comorbidity index is a combined diabetes classification. Previous work has showed there is a large majority (approximately 96\%) with type 2 diabetes and a minority (approximately 4\%) with type 1 diabetes \cite{rafferty2021retrospective, PHW2023DiabetesPrevalence}. Cluster 10 could represent a classic metabolic syndrome profile (diabetes, obesity, hypertension) with relatively low short-term mortality. In contrast, Cluster 32 could reflect an ``end-organ damage'' phenotype, where diabetes is associated with or leads to renal failure, heart failure, and peripheral vascular disease, and is associated with higher mortality. Similar findings have been reported previously \cite{van2011chronic}. Several prior studies identified metabolic and cardiovascular disease clusters \cite{cornell2008multimorbidity, freund2012patterns, garcia2012comorbidity, holden2011patterns, john2003patterns, kirchberger2012patterns, newcomer2011identifying}.

The depression-containing clusters may further illustrate the model’s ability to capture clinically meaningful heterogeneity within a single diagnostic category. Depression in healthcare data represents a broad spectrum of illness, ranging from isolated affective disorders managed in primary care to severe psychiatric disease, and is associated with conditions such as cancer and COPD.
Clusters characterised by high probabilities of depression differed in both the accompanying comorbidities and mortality risk. For example, Cluster 12 appeared to represent predominantly depression with relatively limited physical comorbidity (aside from moderate rates of COPD) and lower mortality, consistent with individuals managed effectively and with mild physical frailty. In contrast, other depression-containing clusters were characterised by substantial co-occurrence with severe physical disease. Cluster 29 showed high rates of depression alongside liver disease, fluid and electrolyte disorders, and other neurological disorders, and was associated with substantially higher mortality. Cluster 26 may highlight a phenotype in which depression co-occurred with cancer, including metastatic disease, and was associated with particularly poor outcomes. Importantly, the relationship between depression and cancer in this cluster is likely bidirectional: in some individuals depression may precede cancer, while in others depression may arise as a consequence of cancer diagnosis, symptom burden, or treatment. Clinically, these patterns are plausible and may reflect the interplay of psychiatric and physical disease. Prior studies also found similar clusters \cite{cornell2008multimorbidity, garcia2012comorbidity, john2003patterns, kirchberger2012patterns}.

We note that, while SVI has proven to be fast and efficient in this specific problem, the method does not sample directly from the model posterior distribution and provides no guarantees that the variational distribution will be a good approximation to the posterior in the general case. It might be necessary for a future workflow utilising SVI in population scale Bayesian modelling to perform a simulation study where SVI can be readily compared with MCMC on smaller datasets as illustrated in this paper, with the final results applied to the large-scale data using SVI. While this is good practice it adds overhead and complexity to modelling projects with large scale data. It is also unclear how to approach the problem if the variational distribution does not appear to be a good approximation to the posterior in a simulation study.  

The flexible framework described in this paper could be further developed in several ways. Any mixture model that features a Dirichlet process may be used as the mixture component. Gaussian and discrete mixtures are the most commonly used and well understood models of this class (for example, the PReMiuM package supports both Gaussian and discrete mixtures \cite{JSSv064i07}), but there are other options depending on the mixture data one wishes to cluster on. For example, Dirichlet process mixture models for count data are well developed \cite{canale2011bayesian}. Furthermore, one may choose to replace the response component of the model with a time-to-event model (including the possibility of competing risks) to understand how the covariates modifying the time to an outcome event differ depending on cluster membership. It may also be of use to develop a hierarchical BPR model in order to capture multiple levels of clustering within disease data. This can be achieved by constructing a Dirichlet process mixture model of Dirichlet processes, with each of the nested Dirichlet processes drawing from a common distribution that reflects the underlying data used to fit the model \cite{teh2006heierarchical}.

We have shown that SVI allows for extremely efficient and scalable inference. It would be of interest to explore how model performance and efficiency is affected by further increases to the dimensionality of the data. For example, modelling clusters of medicines use conditioned on an outcome such as medications related harm. 

\section{Conclusion}
The aim of this work was to develop and apply BPR to a population-scale dataset in a way that is computationally feasible for standard computing hardware. SVI was used to fit a BPR model written using numpyro, a PPL for the python language, on a dataset of millions of individuals demonstrating such datasets can be analysed with this approach. Using simulation studies, results obtained from SVI model fits were found to be appropriate approximations to results derived using MCMC. In an application to routine healthcare data in Wales, UK, clinically plausible disease clusters were identified with the additional advantage that clusters were conditioned on both an outcome of interest and adjusted for covariates. BPR is a useful modelling approach  for clustering data conditioning on  an outcome, and this works demonstrates how this method can be applied to large-scale datasets in order to develop new health/disease insights.

\section{Acknowledgements}

This work was supported by Health Data Research UK (HDRUK2023.0030), which is funded by UK Research and Innovation, the Medical Research Council, the British Heart Foundation, Cancer Research UK, the National Institute for Health and Care Research, the Economic and Social Research Council, the Engineering and Physical Sciences Research Council, Health and Care Research Wales, Health and Social Care Research and Development Division (Public Health Agency, Northern Ireland), Chief Scientist Office of the Scottish Government Health and Social Care Directorates.  Specifically, this work was conducted as part of the HDR-UK Medicines in Acute and Chronic Care Driver programme. RKO is supported by a Health and Care Research Wales - NIHR Advanced Fellowship (HCRW NIHR FS(A)-2023b-RO/NIHR303628) and HCRW Senior Research Leaders Award (SRL-25-019).

The authors thank Kevin Shaw for support during the simulation study and Du Phan, Martin Jankowiak and the members of the numpyro community for their invaluable help during this project. 
This study makes use of anonymised data held in the Secure Anonymised Information Linkage (SAIL) Databank. We would like to acknowledge all the patients and data providers who make anonymised data available for research.

\section{Declarations of interest}

JR declares he has no conflicts of interest.

KRA is a member of the National Institute for Health and Care Excellence (NICE) Diagnostics Advisory Committee, the NICE Decision and Technical Support Units, and is a National Institute for Health Research (NIHR) Senior Investigator Emeritus [NF-SI-0512-10159]. He has served as a paid consultant, providing unrelated methodological and strategic advice, to the pharmaceutical and life sciences industry generally, as well as to DHSC/NICE, and has received unrelated research funding from Association of the British Pharmaceutical Industry (ABPI), European Federation of Pharmaceutical Industries \& Associations (EFPIA), Pfizer, Sanofi and Swiss Precision Diagnostics/Clearblue. He has also received course fees from ABPI and the University of Bristol, and is a Partner and Director of Visible Analytics Limited, a health technology assessment consultancy company.

MP currently receives partnership funding, paid to the University of Liverpool, for the MRC Medicines Development Fellowship Scheme (co-funded by MRC and GSK, AZ, Optum and Hammersmith Medicines Research). He has developed an HLA genotyping panel with MC Diagnostics but does not benefit financially from this. He is part of the IMI Consortium ARDAT (www.ardat.org); none of these funding sources have been used for the current article.

MD is a member of the NICE Breast Faculty and has received fees for advisory board participation and speaker engagements from the pharmaceutical industry.

RKO is a member of the National Institute for Health and Care Excellence (NICE) Technology Appraisal Committee, member of the NICE Decision Support Unit (DSU), and associate member of the NICE Technical Support Unit (TSU). She has served as a paid consultant to the pharmaceutical industry, providing unrelated methodological advice. She reports teaching fees from the Association of British Pharmaceutical Industry (ABPI) and the University of Bristol.

\section{Author contributions}
Conceptualisation of the study: JR, RKO; Data curation: JR, RKO; Analysis: JR; Drafting of the paper: JR, RKO; Review, editing and final approval of the manuscript: JR, KRA, MP, MD, RKO; Funding acquisition: MP, RKO.

\section{Data availability statement}
Access to the data used for all analyses, figures and tables are available to the research community upon approval by the SAIL Databank (https://saildatabank.com/data/apply-to-work-with-the-data/). All proposals to use data held within the SAIL Databank are subject to review and approval by an independent Information Governance Review Panel (IGRP).

\bibliographystyle{unsrt} 
\bibliography{references} 

\pagebreak

\begin{appendices}
\section*{Supplement}
\renewcommand\thefigure{S.\arabic{figure}} 
\renewcommand\thetable{S.\arabic{table}} 
\setcounter{figure}{0}  
\setcounter{table}{0}  
\setcounter{section}{0}
\renewcommand{\thesection}{S.\arabic{section}}

\section{Subgroup analysis - Age Stratification}
\label{sec:age_strat}
\begin{figure}[H]
    \centering
    \includegraphics[width=0.49\linewidth]{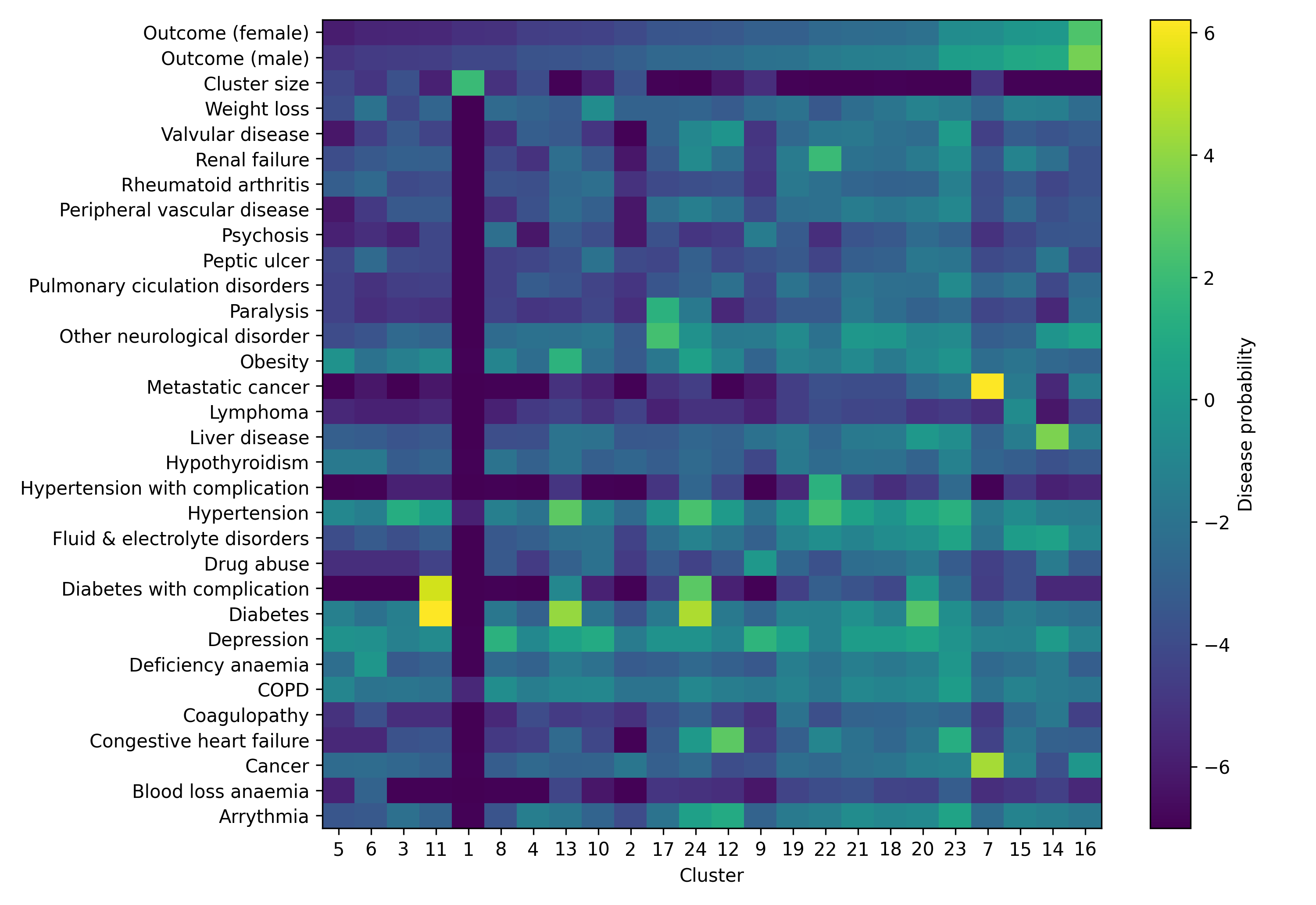}
    \includegraphics[width=0.49\linewidth]{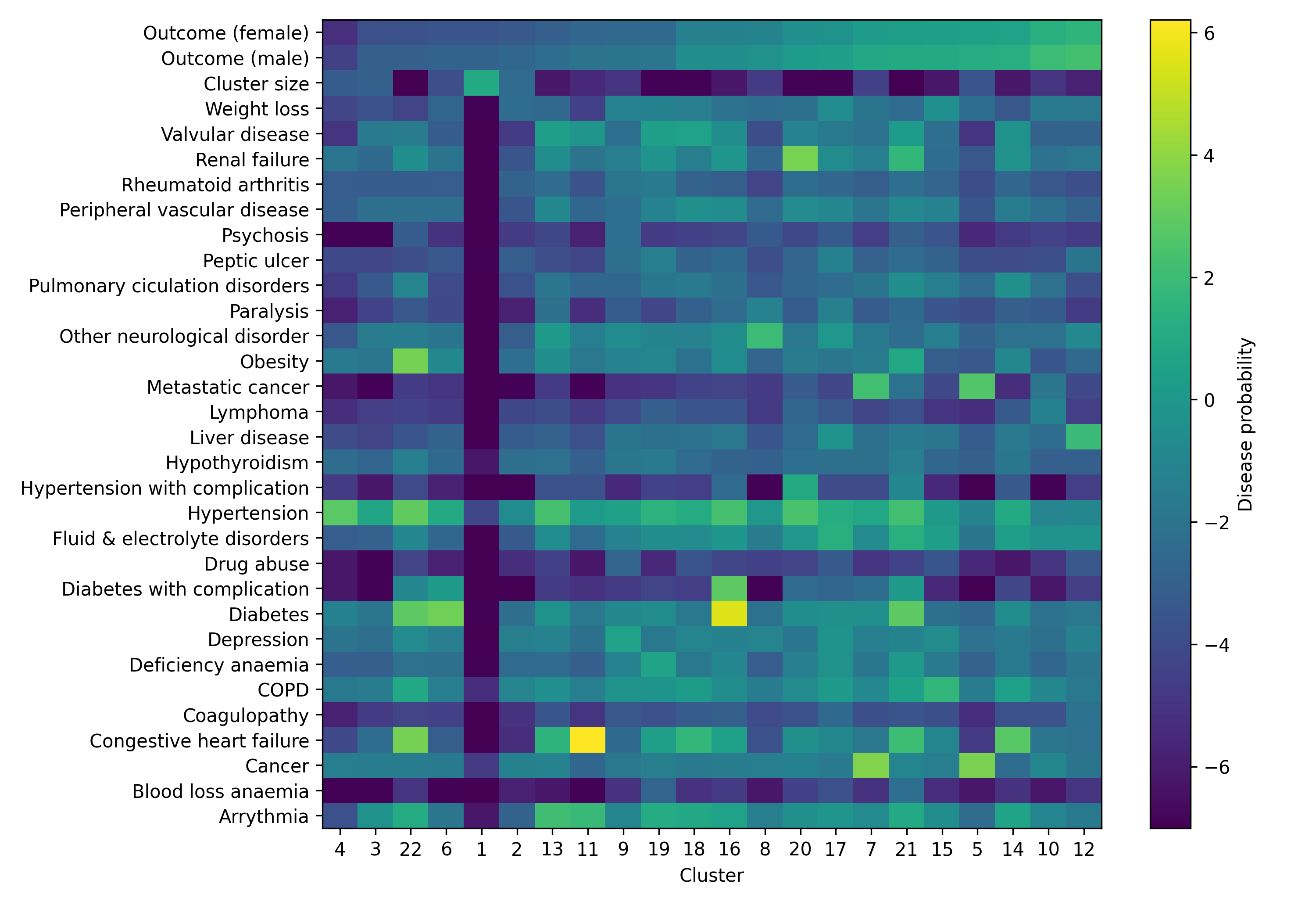}\\
    \includegraphics[width=0.49\linewidth]{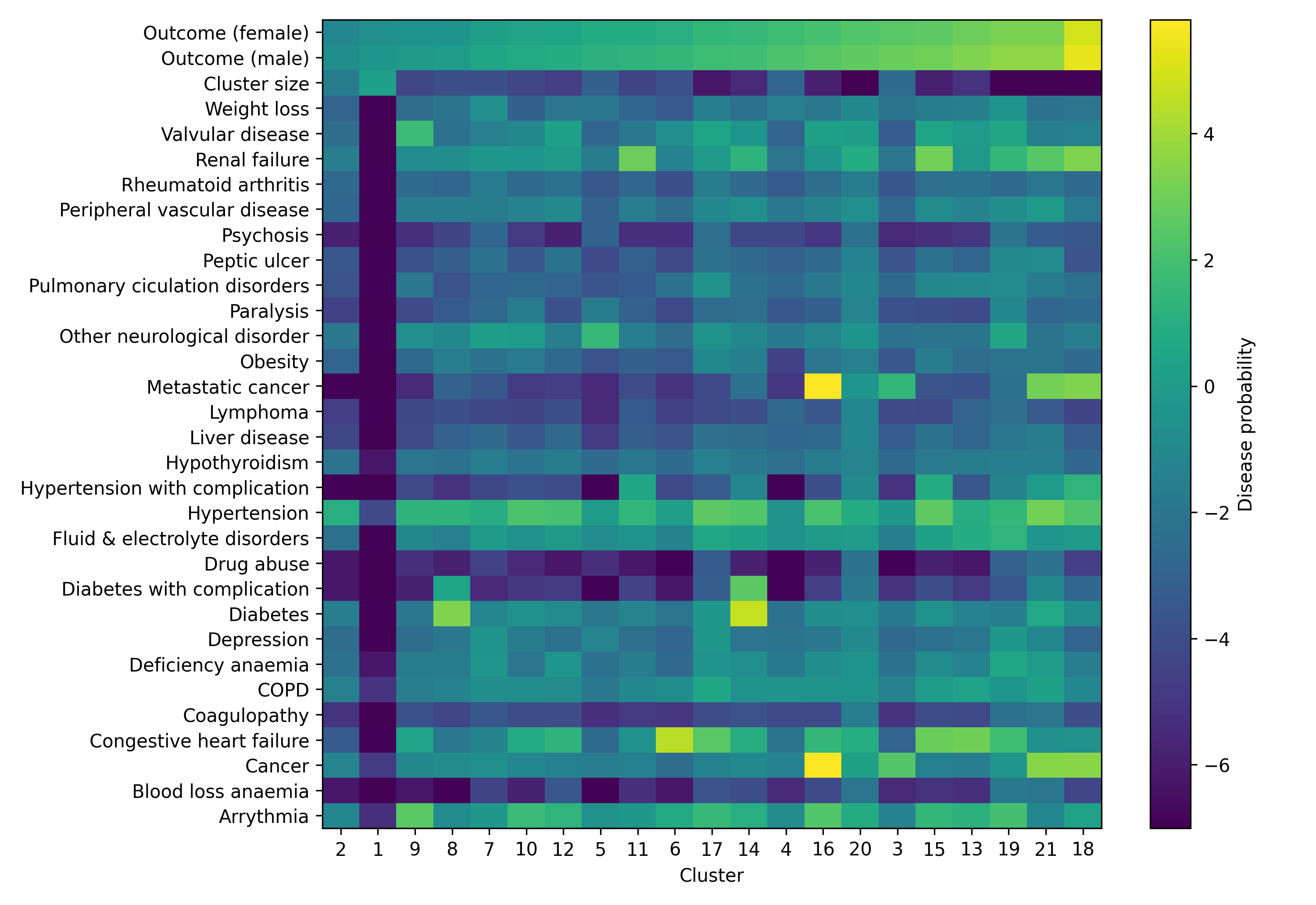}
    \includegraphics[width=0.49\linewidth]{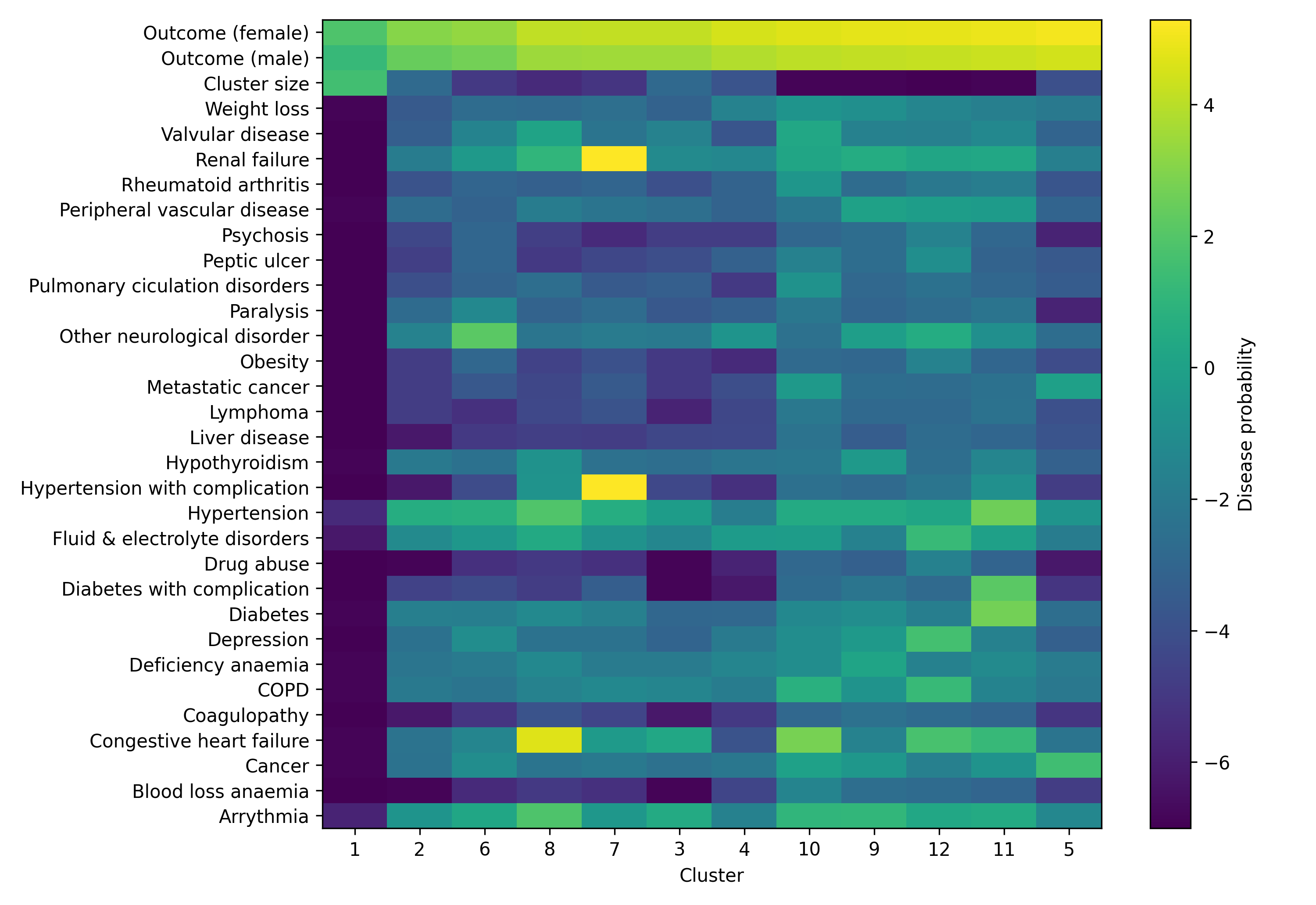}

    \caption{A heatmap showing point estimates for the log-odds of disease and the outcome for each cluster, stratified by age at cohort entry. Top left: $20 < \mathrm{age} \leq 40$ years. Top right: $40 < \mathrm{age} \leq 60$ years. Lower left: $60 < \mathrm{age} \leq 80$ years. Lower right: $80 < \mathrm{age} \leq 100$ years. The outcome is calculated at the mean age each stratum and living in the third deprivation quintile.}
    \label{fig:heatmap_age_strat}
\end{figure}

Results from the age stratified subgroup analysis are shown in Table \ref{tab:coefficients_age_strat} and Figure \ref{fig:heatmap_age_strat}. It was evident that there was always a cluster or clusters where the outcome probability was high, capturing groups at higher risk of mortality at all ages. We note that more clusters with high probability of the outcome and fewer clusters with smaller probability of the outcome feature in older age strata. Diseases associated with older age such as metastatic cancer, congestive heart failure and hypertension with complication appear in in clusters in the younger age strata analysis in fewer, smaller clusters compared to the older age strata. The oldest age stratum, 80 - 100 years was different to the other strata in that the probability of mortality was high for all clusters, there were fewer clusters and the occurrence of some common diseases associated with older age, for example metastatic cancer, were reduced. 

\begin{table}[H]
\centering
\begin{tabular}{lll}
\toprule
Age 20 - 40 years & Coefficient estimate (95\% CrI) & Odds Ratio (95\% CrI) \\
\midrule
Age & 0.043 (0.039, 0.047) & 1.04 (1.04, 1.05) \\
Female sex & -0.952 (-0.996, -0.909) & 0.39 (0.37, 0.40) \\
WIMD Quintile = 1 & 0.391 (0.336, 0.45) & 1.48 (1.40, 1.57) \\
WIMD Quintile = 2 & 0.161 (0.101, 0.209) & 1.17 (1.11, 1.23) \\
WIMD Quintile = 4 & -0.174 (-0.236, -0.116) & 0.84 (0.79, 0.89) \\
WIMD Quintile = 5 & -0.441 (-0.508, -0.377) & 0.64 (0.60, 0.69) \\
\bottomrule
Age 40 - 60 years & Coefficient estimate (95\% CrI) & Odds Ratio (95\% CrI) \\
\midrule
Age & 0.101 (0.099, 0.103) & 1.11 (1.10, 1.11) \\
Female sex & -0.682 (-0.706, -0.657) & 0.51 (0.49, 0.52) \\
WIMD Quintile = 1 & 0.474 (0.44, 0.516) & 1.61 (1.55, 1.68) \\
WIMD Quintile = 2 & 0.227 (0.194, 0.263) & 1.25 (1.21, 1.30) \\
WIMD Quintile = 4 & -0.133 (-0.176, -0.092) & 0.88 (0.84, 0.91) \\
WIMD Quintile = 5 & -0.379 (-0.42, -0.341) & 0.68 (0.66, 0.71) \\
\bottomrule
Age 60 - 80 years & Coefficient estimate (95\% CrI) & Odds Ratio (95\% CrI) \\
\midrule
Age & 0.174 (0.172, 0.177) & 1.19 (1.19, 1.19) \\
Female sex & -0.385 (-0.407, -0.363) & 0.68 (0.67, 0.70) \\
WIMD Quintile = 1 & 0.284 (0.247, 0.323) & 1.33 (1.28, 1.38) \\
WIMD Quintile = 2 & 0.178 (0.146, 0.213) & 1.19 (1.16, 1.24) \\
WIMD Quintile = 4 & -0.128 (-0.17, -0.087) & 0.88 (0.84, 0.92) \\
WIMD Quintile = 5 & -0.283 (-0.326, -0.247) & 0.75 (0.72, 0.78) \\
\bottomrule
Age 80 - 100 years & Coefficient estimate (95\% CrI) & Odds Ratio (95\% CrI) \\
\midrule
Age & 0.007 (0.001, 0.015) & 1.01 (1.00, 1.02) \\
Female sex & 0.623 (0.572, 0.679) & 1.86 (1.77, 1.97) \\
WIMD Quintile = 1 & -0.536 (-0.616, -0.461) & 0.59 (0.54, 0.63) \\
WIMD Quintile = 2 & -0.153 (-0.246, -0.075) & 0.86 (0.78, 0.93) \\
WIMD Quintile = 4 & -0.051 (-0.129, 0.025) & 0.95 (0.88, 1.03) \\
WIMD Quintile = 5 & -0.123 (-0.208, -0.042) & 0.88 (0.81, 0.96) \\
\bottomrule
\end{tabular}
    \caption{Response component model coefficients for the BPR model fit to routine healthcare data, with the cohort stratified by age at cohort entry.}
    \label{tab:coefficients_age_strat}
\end{table}

\section{Cluster probabilities}
\label{sec:cluster_proba}
\begin{table}[H]
\centering
\begin{tabular}{ll}
\toprule
 & Cluster probability (95\% CrI) \\
\midrule
Cluster 1 & 0.783, (0.782, 0.784) \\
Cluster 2 & 0.041, (0.04, 0.042) \\
Cluster 3 & 0.024, (0.023, 0.026) \\
Cluster 4 & 0.018, (0.017, 0.018) \\
Cluster 5 & 0.012, (0.011, 0.013) \\
Cluster 6 & 0.011, (0.01, 0.011) \\
Cluster 7 & 0.011, (0.01, 0.012) \\
Cluster 8 & 0.011, (0.01, 0.011) \\
Cluster 9 & 0.01, (0.009, 0.012) \\
Cluster 10 & 0.009, (0.009, 0.01) \\
Cluster 11 & 0.008, (0.007, 0.008) \\
Cluster 12 & 0.008, (0.007, 0.009) \\
Cluster 13 & 0.007, (0.007, 0.008) \\
Cluster 14 & 0.006, (0.006, 0.006) \\
Cluster 15 & 0.005, (0.005, 0.006) \\
Cluster 16 & 0.004, (0.004, 0.004) \\
Cluster 17 & 0.004, (0.003, 0.004) \\
Cluster 18 & 0.004, (0.004, 0.004) \\
Cluster 19 & 0.003, (0.003, 0.003) \\
Cluster 20 & 0.003, (0.003, 0.004) \\
Cluster 21 & 0.003, (0.003, 0.004) \\
Cluster 22 & 0.002, (0.002, 0.002) \\
Cluster 23 & 0.002, (0.001, 0.002) \\
Cluster 24 & 0.002, (0.001, 0.002) \\
Cluster 25 & 0.002, (0.002, 0.002) \\
Cluster 26 & 0.002, (0.001, 0.002) \\
Cluster 27 & 0.001, (0.001, 0.001) \\
Cluster 28 & 0.001, (0.001, 0.001) \\
Cluster 29 & 0.001, (0.001, 0.001) \\
Cluster 30 & 0.001, (0.001, 0.002) \\
Cluster 31 & 0.001, (0.001, 0.001) \\
Cluster 32 & 0.001, (0.001, 0.001) \\
Cluster 33 & 0.0, (0.0, 0.0) \\
\bottomrule
\end{tabular}
\caption{Cluster probabilities from the BPR model fit to routine healthcare data.}
\label{tab:cluster_proba} 
\end{table}

\section{Linear scale heatmap}
\label{sec:lin_scale}
\begin{figure}[H]
    \centering
    \includegraphics[width=0.8\columnwidth]{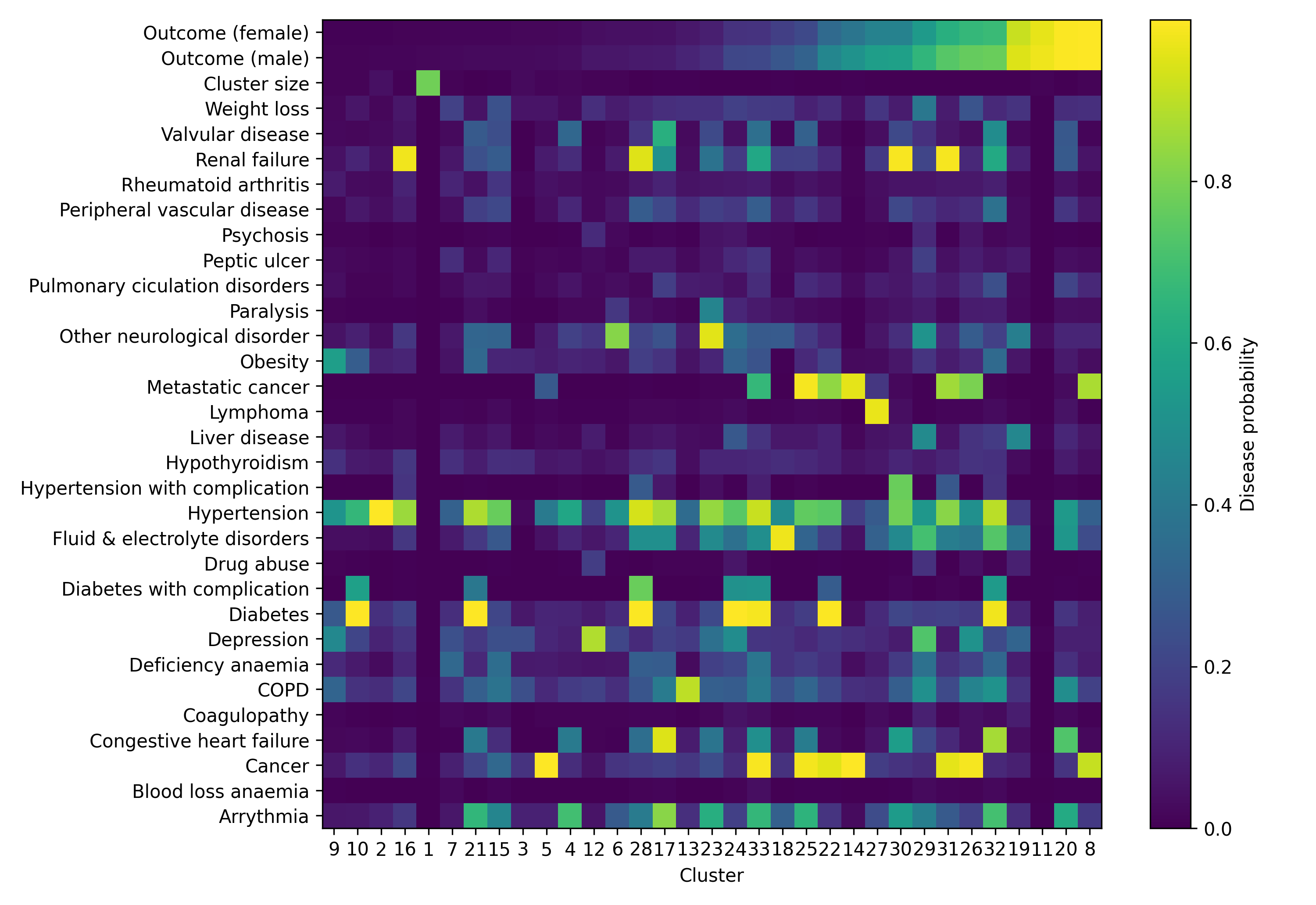}
    \caption{A heatmap showing point estimates for the probability of disease and the outcome for each cluster.}
\end{figure}

\end{appendices}
\end{document}